\documentclass{amsart}

\usepackage{amscd}
\usepackage{amssymb}

\usepackage{graphicx}
\usepackage{sidecap}
\usepackage{subfig}
\usepackage{url}

\theoremstyle{plain}
\newtheorem*{thm*}{Theorem}
\newtheorem{thm}[equation]{Theorem}

\numberwithin{equation}{section}

\DeclareMathOperator{\GL}{GL}

\newcommand{\Z}{\mathbb{Z}}

\newcommand{\E}{\Eg8}
\newcommand{\Eg}[1]{{{\mathsf{E}}_{#1}}}

\newcommand{\R}{\mathbb{R}}
\newcommand{\C}{\mathbb{C}}

\newcommand{\g}{\mathfrak{g}}
\newcommand{\h}{\mathfrak{h}}

\newcommand{\su}{\mathfrak{su}}
\newcommand{\so}{\mathfrak{so}}

\newcommand{\ra}{\rightarrow}
\newcommand{\ot}{\otimes}

\newcommand{\vx}{\vec{x}}

\DeclareMathOperator{\SU}{SU}

\newcommand{\s}{\sigma}

\renewcommand{\O}{\Omega}

\newcommand{\credit}[1]{{\small{(#1)}}}

\begin{document}

\title{Did a 1-dimensional magnet detect a 248-dimensional Lie algebra?}

\author{David Borthwick}
\email{davidb \text{at} mathcs.emory.edu}

\author{Skip Garibaldi}
\email{skip \text{at} mathcs.emory.edu}


\begin{abstract}
About a year ago, a team of physicists reported in \emph{Science} that they had observed ``evidence for $\E$ symmetry'' in the laboratory.  This expository article is aimed at mathematicians and explains the chain of reasoning connecting measurements on a quasi-1-dimensional magnet with a 248-dimensional Lie algebra.
\end{abstract}

\thanks{DB was partially supported by the NSF under grant DMS-0901937.  We are grateful to Richard Borcherds, Radu Coldea, Jacques Distler, Giuseppe Mussardo, the referees, and others for their helpful remarks.  We thank Bert Kostant and Richard Borcherds for bringing this topic to our attention.}

\maketitle

You may have heard some of the buzz spawned by the recent paper \cite{Coldea} in \emph{Science}.  That paper described a neutron scattering experiment involving a quasi-one-dimensional cobalt niobate magnet, and led to rumors that $\E$ had been detected ``in nature".  This is fascinating, because $\E$ is a mathematical celebrity and because such a detection seems impossible: it is hard for us to imagine a realistic experiment that could directly observe a 248-dimensional object like $\E$.

The connection between the cobalt niobate experiment and $\E$ is as follows.  Around 1990, physicist Alexander Zamolodchikov and others studied perturbed conformal field theories in general; one particular application of this was a theoretical model describing a 1-dimensional magnet subjected to two magnetic fields.  This model makes some numerical predictions which were tested in the cobalt niobate experiment, and the results were as predicted by the model.  As the model involves $\E$ (in a way we will make precise in \S\ref{ATFT}), one can say that the experiment provides evidence for ``$\E$ symmetry".  No one is claiming to have directly observed $\E$.

Our purpose here is to fill in some of the details omitted in the previous paragraph.  We should explain that we are writing as journalists rather than mathematicians here, and we are not physicists.  We will give pointers to the physics literature so that the adventurous reader can go directly to the words  of the experts for complete details.

\section{The Ising model} \label{Ising.sec}

The article in \emph{Science} describes an experiment involving the magnetic material 
cobalt niobate (CoNb$_2$O$_6$).
\begin{figure}[hbt]
\includegraphics{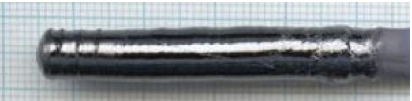}
\caption{Photograph of an artificially-grown single crystal of CoNb$_2$O$_6$.  The experiment involved a 2-centimeter-long piece of this crystal, weighing about 8 grams.
\credit{Image courtesy of Radu Coldea.}} \label{magnet.fig}
\end{figure}
The material was chosen because the internal crystal structure is such that magnetic Co$^{2+}$ ions are arranged into long chains running along one of the crystal's axes,
and this could give rise to 1-dimensional magnetic behavior.\footnote{Two additional practical constraints led to this choice of material: (1) large, high-quality single crystals of it can be grown as depicted in Figure \ref{magnet.fig}, and (2) the strength of the
magnetic interactions between the Co$^{2+}$ spins is
low enough that the
quantum critical point corresponding to $g_x=1$ in (\ref{quant.H}) can be matched by magnetic fields currently achievable in the laboratory.}  
In particular, physicists expected that this material would provide a realization of the famous
Ising model, which we now describe briefly. 

The term \emph{Ising model} refers generically to the original, classical model.\footnote{For more details on this model, see for example 
\cite{McCoyWu:book} or \cite[Chap.~12]{DMS}.}
This simple model for magnetic interactions was suggested by W.~Lenz as a thesis problem
for his student E.~Ising, whose thesis appeared in Hamburg in 1922 \cite{Ising}.  The classical  form 
of the model is built on a square, periodic, $n$-dimensional lattice, with the periods sufficiently large
that the periodic boundary conditions don't play a significant role in the physics.
Each site $j$ is assigned a spin $\sigma_j = \pm 1$, interpreted as the projection of the 
spin onto some preferential axis.   The energy of a given configuration of spins is
\begin{equation}\label{class.H}
H =  - J \sum_{\langle i,j\rangle} \sigma_i \sigma_j, 
\end{equation}
where $J$ is a constant and the sum ranges over pairs $\langle i,j\rangle$ of nearest-neighbor 
sites.  This Hamiltonian gives rise to a statistical ensemble of states that is used to model
the thermodynamic properties of actual magnetic materials.  This essentially amounts to
a probability distribution on the set of spin configurations, with each configuration weighted
by $e^{-kH/T}$ (the Boltzmann distribution), where $k$ is constant and $T$ is the temperature.
The assumption of this distribution makes the various physical quantities, such as
individual spins, average energy, magnetization, etc., into random variables.  
For $J>0$, spins at neighboring sites tend to align in the same direction; this behavior is called 
\emph{ferromagnetic}, because this is what happens with iron.

To describe the cobalt niobate experiment we actually want the 
quantum spin chain version of the Ising model.  In this quantum model, each site in a 1-dimensional 
(finite periodic) chain is assigned a 2-dimensional complex Hilbert
space.  The Pauli spin matrices $S^x$, $S^y$, and $S^z$ act
on each of these vector spaces as spin observables, meaning they are self-adjoint operators
whose eigenstates correspond to states of particular spin.  For example,
the $\pm 1$ eigenvectors of $S^z$ correspond to up and down spins along the $z$-axis.
A general spin state is a unit vector in the 2-dimensional Hilbert space, which could be viewed as a 
superposition of up and down spin states, if we use those eigenvectors as a basis.   

The Hamiltonian operator for the standard 1-dimensional quantum Ising model is given by
\begin{equation}\label{quant.H}
\hat{H} = - K \sum_j \Bigl[ S^z_j S^z_{j+1} + g_x S^x_j \Bigr].
\end{equation}
In the quantum statistical ensemble one assigns probabilities to 
the eigenvectors of $\hat{H}$ weighted by the corresponding energy eigenvalues. 
This then defines, via the Boltzmann distribution again, a probability distribution on the 
unit ball in the total Hilbert space of the system. 

Just as in the classical case, physical quantities
become random variables with distributions that depend on the temperature and constants
$K$ and $g_x$.  It is by means of these distributions that the model makes 
predictions about the interrelationships of these quantities. 

The first term in the Hamiltonian (\ref{quant.H}) has a ferromagnetic effect (assuming $K>0$), just 
as in the classical case.  That is, it causes spins of adjacent sites to align with each other
along the $z$-axis, which we will refer to as the \emph{preferential} axis.  
(Experimental physicists might call this the ``easy" axis.)
The second term represents the influence
of an external magnetic field in the $x$-direction, perpendicular to the $z$-direction---we'll refer to this as the \emph{transverse} axis.  The effect of the second term is \emph{paramagnetic},
meaning that it encourages the spins to align with the transverse field.

The 1-dimensional quantum Ising spin chain exhibits a phase transition at zero temperature.
The phase transition (also called a critical point) 
is the point of transition between the ferromagnetic regime ($g_x < 1$, where spins tend to align 
along the $z$-axis) and the paramagnetic ($g_x > 1$).  The critical point $(g_x=1)$ is distinguished by
singular behavior of various macroscopic physical quantities, such as
the \emph{correlation length}.  Roughly speaking, this is the average size of the regions in which
the spins are aligned with each other.  

To define correlation length a little more precisely, we consider the
statistical correlation between the $z$-components of spins at two sites separated by a distance 
$r$.   These spins are just random variables whose joint distribution depends on the constants 
$K$ and $g_x$ as well as the separation $r$ and the temperature $T$.   (We are assuming $r$ 
is large compared to the lattice spacing, but small compared to the overall dimensions of the system.)  
For $g_x > 1$ the correlation falls off exponentially as $e^{-r/\xi}$, because spins lined up along the
$x$-axis will be uncorrelated in the $z$-direction; the constant $\xi$ is the correlation length.
In contrast, at the critical point $g_x=1$ and $T=0$, the decay of the spin correlation is given by a power law;
this radical change of behavior corresponds to the divergence of $\xi$.  This phase transition has been observed 
experimentally in a LiHoF$_4$ magnet \cite{BRA}.

One might wonder why an external magnetic field is included by default in the
quantum case but not in the classical case.
The reason for this is a correspondence
between the classical models and the quantum models of one lower dimension.
The quantum model includes a notion of time-evolution of an observable 
according to the Schr\"odinger equation, and the correspondence involves interpreting
one of the classical dimensions as imaginary time in the quantum model.
Under this correspondence, the classical interaction in the spatial directions gives
the quantum ferromagnetic term, while interactions in the imaginary time direction
give the external field term, see \cite[\S2.1.3]{Sachdev} for details.   

Although there are some important differences in the
physical interpretation on each side, the classical-quantum correspondence
allows various calculations to be carried over from one case to the other.
For example, the critical behavior of the 1-dimensional transverse-field quantum
Ising model (\ref{quant.H}) at zero temperature, with the transverse field parameter
tuned to the critical value $g_x=1$, can be ``mapped'' onto equivalent physics for the
classical 2-dimensional Ising model (\ref{class.H}), at a non-zero temperature.
The latter case is the famous phase transition of the 2-dimensional classical model,
which was discovered by Peierls and later solved exactly by Onsager \cite{Onsager}.

\section{Adapting the model to the magnet}\label{adapt.sec}

The actual magnet used in the experiment is not quite modeled by the quantum Ising Hamiltonian \eqref{quant.H}.  
In the ferromagnetic regime ($g_x< 1$), weak couplings between the magnetic chains create an effective magnetic field pointing along the preferential axis \cite{CarrTsvelik}.  The relevant model for the experiment is thus
\begin{equation}\label{Hmag}
\hat{H} = - K \sum_{j} \Bigl[ S^z_j S^z_{j+1} + g_x S^x_j + g_z S^z_j\Bigr],
\end{equation}
which is just \eqref{quant.H} with an additional term $g_z S^z_j$ representing this internal magnetic field.

\begin{figure}[hbt]
\centering
\subfloat[Experimental data at 40mK]{\includegraphics[width=0.42\textwidth]{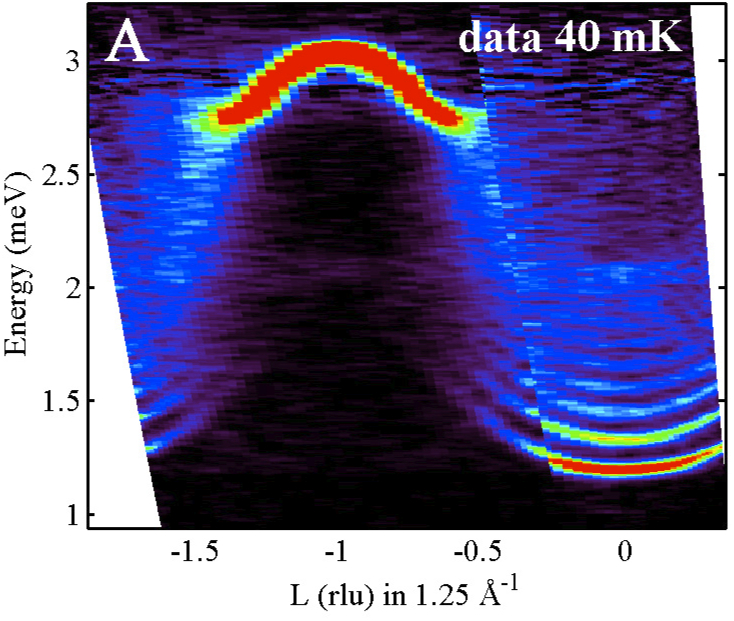}}
\qquad
\subfloat[Calculated]{\includegraphics[width=0.42\textwidth]{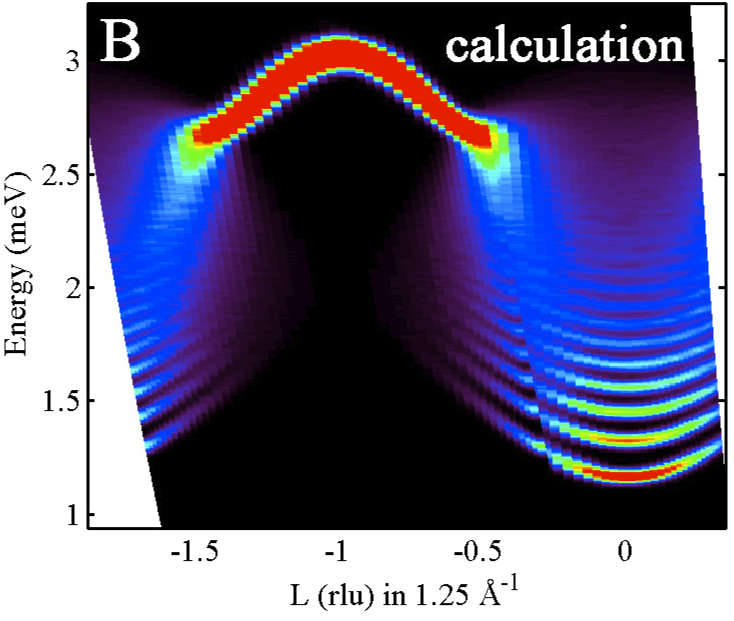}}
\caption{Comparison of excitations under no external magnetic field: experimental
(left) versus predictions based on the 1-dimensional model (right).  \credit{Figure adapted from \cite{Coldea}.}} \label{zerofield}
\end{figure}

The first phase of the cobalt niobate experiment tested the appropriateness of \eqref{Hmag}
as a model for the magnetic dynamics in the absence of an external magnetic field, i.e., with $g_x = 0$.
The experimental
evidence does support the claim that this 3-dimensional object is behaving as a 1-dimensional
magnetic system.  For example, Figure \ref{zerofield} shows a comparison of the experimental 
excitation energies (as a function of wave vector) to theoretical predictions from the 1-dimensional model.  The presence of a sequence of well-defined and closely spaced 
energy levels, as shown in these pictures, is predicted only in dimension one.

\section{What is $\E$?}

\begin{figure}[hbt]
\centering
\subfloat[The popular picture]{\label{gosset.pop}\includegraphics[width=0.42\textwidth]{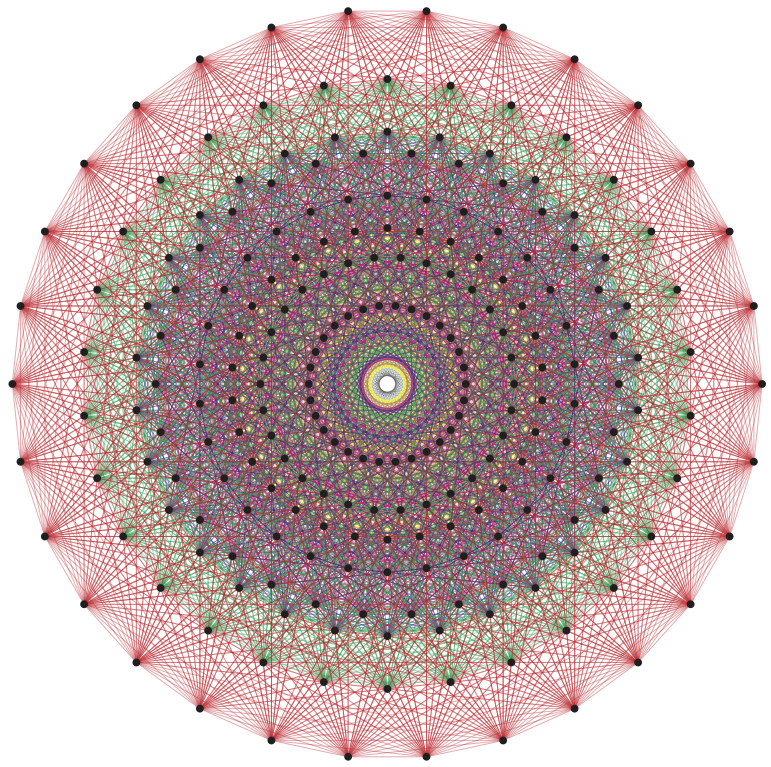}}
\qquad
\subfloat[Vertices only]{\label{gosset.actual}\includegraphics[width=0.42\textwidth]{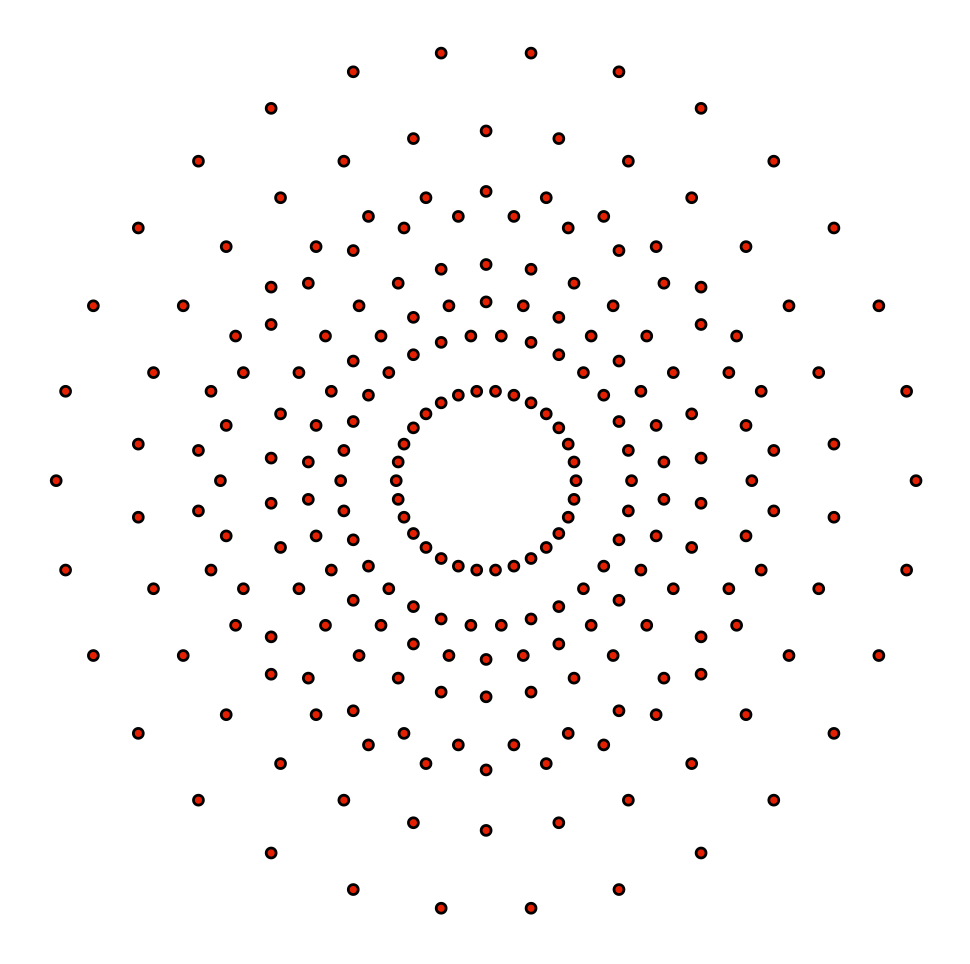}}
\caption{The left panel (A) is the picture of $\E$ that one finds in the popular press.  Deleting some edges leaves you with the frontispiece of \cite{Cox:RCP}.  \credit{Image courtesy of John Stembridge \cite{Stembridge}.} The right panel (B) is the same picture with the edges removed; it is the image of the root system of $\E$ in a Coxeter plane.} \label{gosset.fig}
\end{figure}

Before we explain how the rather simple quantum Ising model from the previous sections leads to a theory involving $\E$, we had better nail down what  it means to speak of ``$\E$".  It's an ambiguous term, with at least the following six common meanings:
\begin{enumerate}
\item The \emph{root system} of type $\E$.  This is a collection of 240 points, called \emph{roots}, in $\R^8$.  The usual publicity photo for $\E$ (reproduced in Figure \ref{gosset.pop}) is the orthogonal projection of the root system onto a copy of $\R^2$ in $\R^8$.  

\item The $\E$ \emph{lattice}, which is the subgroup of $\R^8$ (additively) generated by the root system.  

\item \label{list.C} A \emph{complex Lie group}---in particular, a closed subgroup of $\GL_{248}(\C)$---that is simple and 248-dimensional.  
\end{enumerate}

There are also three simple \emph{real} Lie groups---meaning in particular that they are closed subgroups of $\GL_{248}(\R)$---whose complexification is the complex Lie group from \eqref{list.C}.  (The fact that there are exactly three is part of Elie Cartan's classification of simple real Lie groups; see \cite[\S{II.4.5}]{SeCG} for an outline of a modern proof.) They are:
\begin{enumerate}
\setcounter{enumi}{3}
\item The \emph{split} real $\E$.  This is the form of $\E$ that one can define easily over any field or even over an arbitrary scheme.  Its Killing form has signature 8.  
\item The \emph{compact} real $\E$, which is the unique largest subgroup of the complex $\E$ that is compact as a topological space.  Its Killing form has signature $-248$.  
\item The remaining real form of $\E$ is sometimes called ``quaternionic".  Its Killing form has signature $-24$.
\end{enumerate}

In physics, the split real $\E$ appears in supergravity \cite{MarcusSchwarz} and the compact real $\E$ appears in heterotic string theory \cite{GHMR}.  
These two appearances in physics, however, are purely theoretical; 
the models in which they are appear are not yet subject to experiment.  
It is the compact real $\E$ (or, more precisely, the associated Lie algebra)
that appears in the context of the cobalt niobate experiment,
making this the first actual experiment to detect a phenomenon that could be modeled using 
$\E$.

There have also been two recent frenzies in the popular press concerning $\E$.  
One concerned the computation of the Kazhdan-Lusztig-Vogan polynomials which you can read about in the prizewinning paper \cite{Vogan:atlas}; that work involved the split real $\E$.  The other frenzy was sparked by the manuscript \cite{Lisi}.  The $\E$ referred to in \cite{Lisi} is clearly meant to be one of the real forms, but the manuscript contains too many contradictory statements to be sure which one\footnote{There are three places in \cite{Lisi} where a particular form of $\E$ might be specified.  At the top of page 18 is a form containing a product of the non-split, non-compact form of $\mathsf{F}_4$ and the compact $\mathsf{G}_2$; therefore it is the split real $\E$ by \cite[p.~118]{Jac:ex} or \cite[\S3]{GS:deg5}.  The form of $\E$ described in the middle of page 21 is supposed to contain a copy of $\so(7,1) \oplus \so(8)$, but there is no such real form of $\E$.  Finally, on page 29, the quaternionic $\E$ is mentioned in the text.}, and in any case the whole idea has serious difficulties as explained in \cite{DiG}.

\subsection*{Groups versus algebras}
Throughout this note we conflate a real Lie group $G$, which is a manifold, with its Lie algebra $\g$, which is the tangent space to $G$ at the identity and is a real vector space endowed with a nonassociative multiplication.  This identification is essentially harmless and is standard in physics.   Even when physicists discuss symmetry
``groups'', they are frequently interested in symmetries that hold only in a local sense, 
and so the Lie algebra is actually the more relevant object.

\subsection*{Real versus complex}
Moreover, physicists typically compute within the complexification $\g \ot \C$ of $\g$.  This is the complex vector space with elements of the form $sx + ity$ for $s, t \in \R$ and $x, y \in \g$, where complex conjugation acts via $sx + ity \mapsto sx - ity$.  Note that one can recover $\g$ as the subspace of elements fixed by complex conjugation.  Therefore, morally speaking, working with the $\R$-algebra $\g$ (as mathematicians often do) amounts to the same as working with $\g \ot \C$ together with complex conjugation (as physicists do).  This is an example of the general theory of Galois descent as outlined in, e.g., \cite[\S{X.2}]{Jac:LA} or \cite[\S{X.2}]{SeLF}.

\section{From the Ising model to $\E$}

What possible relevance could a 248-dimensional algebra have for
a discrete one-dimensional statistical physics model?  This is a long and 
interesting story, and we can only give a few highlights here.

As we mentioned above, the 1-dimensional quantum Ising model from \eqref{quant.H}
undergoes a phase transition at zero temperature at the critical value of the 
transverse magnetic field strength.  If the system is close to this critical point,
the correlation length (described in \S\ref{Ising.sec}) will be very large compared to
the lattice spacing, and so we can assume that the discrete spins vary smoothly
across nearby lattice sites.  In this regime we can thus effectively model the system 
using continuous ``field'' variables, i.e., using quantum field theory.
For the 1-dimensional quantum Ising model, the corresponding continuous theory is a quantum 
field theory of free, spinless fermionic particles in 1+1 space-time dimensions.  

To understand what happens as the critical point is approached, one can apply ``scaling'' 
transformations that dilate the macroscopic length scales (e.g.~the 
correlation length) while keeping the microscopic lengths (e.g.~the lattice spacing) unchanged.  
(See, e.g., \cite[\S4.3]{Sachdev} for a more thorough explanation of this.)
The limiting theory at the critical point should then appear as a fixed point for these 
transformations, called the \emph{scaling limit}.
Polyakov famously argued in \cite{Polyakov} that the scaling limit should
be distinguished by invariance with respect to local conformal transformations.  
This paper established the link between the study of phase
transitions and \emph{conformal field theory} (CFT).

In \cite{BPZ:NucB}, Belavin, Polyakov, and Zamolodchikov showed that certain simple CFT's 
called \emph{minimal models} could be solved completely in terms of (and so are determined by) a Hilbert space made 
of a finite number of ``discrete series" (unitary, irreducible) representations of the Virasoro algebra, see 
\cite[Chap.~2]{Henkel:book} or \cite[Chap.~7]{DMS} for more details.   
These representations are characterized by the 
eigenvalue $c$ assigned to the central element, called the \emph{central charge}, 
which can be computed directly
from the scaling limit of the statistical model.  
This works out beautifully in the case of the critical 1-dimensional quantum Ising model:
In that case, the central charge is $c = 1/2$, the minimal model is built from 
the three discrete series representations of the Virasoro algebra with that central charge,
 and this CFT exactly matches the Ising phase transition, see \cite[App.~E]{BPZ:NucB}, \cite[\S7.4.2]{DMS}, or
\cite[\S14.2]{Mussardo} for details.

The discrete series representations mentioned above are described by $c$ and another parameter $h$ which have some relations between them, and there are tight constraints on the possible values of $c$ and $h$ to be unitary \cite{FQS:2D}.  To prove that all of these values of $c$ and $h$ indeed correspond to irreducible unitary representations, one employs the \emph{coset construction} of Goddard, Kent, and Olive, see \cite{GKO:Vir} or \cite[Chap.~18]{DMS}.  This construction produces such representations by restricting representations of an affine Lie algebra, i.e., a central extension of the (infinite-dimensional) loop algebra of a compact Lie algebra $\g$.  Using the coset construction, there are two
ways to obtain the $c=1/2$ minimal model that applies to our zero-field Ising model:
we could use either of the compact Lie algebras $\su(2)$ or $\E$ as the base $\g$ for the
affine Lie algebra \cite[\S18.3, \S18.4.1]{DMS}.  These two algebras are the only choices that lead to 
$c=1/2$ \cite[\S14.2]{Mussardo}.

Of course, the appearance of $\E$ here is somewhat incidental.  The minimal model could be
described purely in terms of Virasoro representations, without reference to either $\su(2)$ or $\E$. 
As we explain below, $\E$ only takes center stage when we
consider a perturbation of the critical Ising model as in \eqref{Hmag}.

\section{Magnetic perturbation and Zamolodchikov's calculation} \label{Zam.sec}

In a 1989 article \cite{Zam:int}, Zamolodchikov investigated the field theory for a model equivalent to the 1-dimensional quantum Ising model \eqref{quant.H}, in the vicinity of the critical point, but perturbed by a small magnetic field directed along the preferential spin axis.  In other words, he considered the field theory model
corresponding to (\ref{Hmag}) with $g_x \approx 1$ and $g_z$ very small.  Note the change of perspective: for Zamolodchikov $g_x$ is fixed and the perturbation consists of a small change in the value of $g_z$.
But in the cobalt niobate 
experiment, this magnetic ``perturbation'' is already built-in---it is the purely internal effect arising from the inter-chain interactions as we described in \S\ref{adapt.sec}.
The experimenters can't control the strength of the internal field, they only vary $g_x$.   
Fortunately, the internal magnetic field $g_z$ turns out to be relatively weak, 
so when the external field $g_x$ is tuned close to the critical value the experimental model matches the
situation considered by Zamolodchidkov.  

The qualitative features of the particle spectrum for the magnetically perturbed Ising model
had been predicted by McCoy and Wu \cite{MW}.  Those earlier calculations show a large number
of stable particles for small $g_x$, with the number decreasing as $g_x$ approaches $1$.  
Zamolodchikov's paper makes some predictions for the masses of these particles at $g_x = 1$.

As we noted above, the $c=1/2$ minimal model is the conformal field theory associated with the 
phase transition of the unperturbed quantum Ising model.  The perturbed field theory is no 
longer a conformal field theory,
but Zamolodchikov found six local integrals of motion for the perturbed field theory 
and conjectured that these were the start of an infinite series.  On this basis, he made the
fundamental conjecture:
\begin{enumerate}
\renewcommand{\theenumi}{{Z\arabic{enumi}}}
\item \label{Z0} The perturbation gives an integrable field theory.
\end{enumerate}
One implication of \eqref{Z0} is that the resulting scattering theory should be ``purely elastic,'' 
meaning that the number of particles and their individual momenta would be conserved asymptotically.  
Zamolodchikov combined this purely elastic scattering assumption with three rather mild assumptions on the 
particle interactions of the theory \cite[p.~4236]{Zam:int}:
\begin{enumerate}
\renewcommand{\theenumi}{{Z\arabic{enumi}}}
\setcounter{enumi}{1}
\item \label{Z1} There are at least 2 particles, say $p_1$ and $p_2$.
\item \label{Z2} Both $p_1$ and $p_2$ appear as bound-state poles on the scattering amplitude
for two $p_1$'s.
\item \label{Z3} The particle $p_1$ appears as a bound-state pole in the scattering amplitude
between $p_1$ and $p_2$.
\end{enumerate}
Assumptions \eqref{Z2} and \eqref{Z3} merely assert that certain coupling constants that
govern the interparticle interactions are non-zero, so they could be viewed as an assumption 
of some minimum level of interaction between the two particles.

The word ``particle" bears some explaining here, because it is being used here in the sense 
of quantum field theory: a stable excitation of the system with distinguishable particle-like features
such as mass and momentum.  However, it is important to note that 
the continuum limit of the Ising model is made to look like a field theory only
through the application of a certain transformation (Jordan-Wigner, see
\cite[\S4.2]{Sachdev}),
that makes ``kink'' states (boundaries between regions of differing spin) 
the basic objects of the theory.
So Zamolodchikov's particles aren't electrons or ions.  The field theory excitations 
presumably correspond to highly complicated aggregate spin states of the original system.
On the statistical physics side the usual term for this kind of excitation is 
\emph{quasiparticle}.  In the experiment these quasiparticles are detected just as ordinary
particles would be, by measuring the reaction to a beam of neutrons.

From the mild assumptions \eqref{Z1}--\eqref{Z3}, he showed that the
simplest purely elastic scattering theory consistent with the integrals of motion contains
8 particles with masses listed in Table \ref{masses}.  (See \cite[\S14.3]{Henkel:book} for more
background on these calculations.)
These predictions were quickly corroborated by computational methods, through numerical diagonalization of the Hamiltonian (\ref{Hmag}), 
see \cite{HenkelSaleur} or \cite{SagdeevZ}.  In Table \ref{masses}, $m_1$ and $m_2$ are the masses of the two original particles $p_1$ and $p_2$.  Note that only the ratios of the masses, such as $m_2/m_1$, are predicted; in the 
discrete model (\ref{Hmag}) the individual masses would depend on the overall length of the lattice, and
in passing to the scaling limit we give up this information.

\begin{table}[htb]
\[
\begin{array}{c@{\ =\ }c@{\ \approx\ }c}
m_2 &2 \cos \frac{\pi}5 m_1& 1.618m_1\\[0.5ex]
  m_3& 2 \cos \frac{\pi}{30} m_1&1.989m_1\\[0.5ex]
m_4 & 2 \cos \frac{\pi}5 \cos \frac{7\pi}{30} m_1&2.405m_1 \\[0.5ex]
m_5 & 4 \cos\frac{\pi}5 \cos \frac{2\pi}{15} m_1 & 2.956m_1\\ [0.5ex]
m_6 & 4 \cos \frac{\pi}5 \cos \frac{\pi}{30} m_1 & 3.218m_1\\[0.5ex]
m_7 & 8 (\cos \frac{\pi}5)^2 \cos \frac{7\pi}{30} m_1 & 3.891m_1\\[0.5ex]
 m_8 & 8 (\cos \frac{\pi}{5})^2 \cos \frac{2\pi}{15} m_1 & 4.783m_1
\end{array}
\]
\caption{The masses of the particles predicted by Zamolodchikov.} \label{masses}
\end{table}

Zamolodchikov's results give some indications of a connection with the algebra or root system $\E$.
The spins of the six integrals of motion he calculated were
$$
s = 1, 7, 11, 13, 17, 19.
$$
The conjecture is that this is the start of a sequence of integrals of motion whose spins include
all values of $s$ relatively prime to $30$.  These numbers are suggestive because 30 is the Coxeter number of $\E$ and the remainders of these numbers modulo 30 are the exponents of $\E$ (see for example \cite{Bou:g4} for a definition of Coxeter number and exponent).  
This was taken as 
a hint that the conjectured integrable field theory could have a model based on $\E$, and in fact
such a connection with $\E$ had already been conjectured by Fateev based on other theoretical 
considerations \cite[p.~4247, 4248]{Zam:int}.

\section{Affine Toda field theory} \label{ATFT}

Soon after Zamolodchikov's first paper appeared, Fateev and Zamolodchikov conjectured 
in \cite{FZ} that if you take a minimal model CFT constructed from a compact Lie algebra $\g$ via the coset construction and perturb it in a particular way, then you obtain the \emph{affine Toda field theory} (ATFT) associated with $\g$, which is an integrable field theory.  This was confirmed in \cite{EguchiYang} and \cite{HollowoodMansfield}.

If you do this with $\g = \E$, you arrive at the conjectured integrable field theory investigated by Zamolodchikov and described in the previous paragraph.  That is, if we take the $\E$ ATFT as a starting point, then the assumptions \eqref{Z0}--\eqref{Z3} become deductions.  This is the essential role of $\E$ in the numerical predictions relevant to the cobalt niobate experiment.  (In the next section, we will explain how the masses that Zamolodchikov found arise naturally in terms of the algebra structure.  But that is just a bonus.)

\subsection*{What is the role of $\E$ in the affine Toda field theory?}  To say the ATFT in question
is ``associated'' with $\E$ leaves open a range of possible interpretations,
so we should spell out precisely what this means. 
The ATFT construction from a compact Lie algebra $\g$ proceeds by choosing a Cartan subalgebra\footnote{It doesn't matter which one you choose, because any one can be mapped to any other via some automorphism of $\g$.} $\h$ in $\g$---it is a real inner product space with inner product the Killing form $(\,,\,)$, and is isomorphic to $\R^8$ in the case $\g = \E$.  Let $\phi$ be a scalar field
in 2-dimensional Minkowski space-time, taking values in $\h$.  Then the Lagrangian density for the affine Toda field theory is
\begin{equation}
\frac12 (\partial_\mu \phi, \partial^\mu \phi) -  (e^{\beta \phi} E e^{-\beta \phi}, \overline{E}),
\end{equation}
where $\beta$ is a coupling constant.  Here $E$ is a regular semisimple element of $\g \ot \C$ that commutes with its complex conjugate $\overline{E}$.  More precisely, for $x \in \h$ a principal regular element, conjugation by $e^{2\pi i x/h}$ with $h$ the Coxeter number of $\g$ gives a $\Z/h$-grading on $\g \ot \C$, and the element $E$ belongs to the $e^{2 \pi i/h}$-eigenspace.
(Said differently, the centralizer of $E$ is a Cartan subalgebra of $\g \ot \C$ in apposition to $\h \ot \C$ in the sense of \cite[p.~1018]{Kost:prin}.)

The structure of $\E$ thus enters into the basic definitions of the fields and their interactions.  
However, $\E$ does not act by symmetries on this set of fields.  

\subsection*{Why is it $\E$ that leads to Zamolodchikov's theory?}
We opened this section by asserting that perturbing a minimal model CFT constructed from $\g$ via the coset construction leads to an ATFT associated with $\g$.  For this association to make sense, the perturbing field is required to have ``conformal dimension"
$2/(h+2)$.  
The two coset models for the Ising model give us two possible perturbation theories.  
Starting from $\su(2)$, which has $h=2$, we could perturb using the field of conformal dimension 
$1/2$, which is the energy.  This perturbation amounts to raising the temperature away 
from zero, which falls within the traditional framework of the Ising model and is well-understood.

The other choice is to start from $\E$, which has $h = 30$, 
and perturb using the field of conformal dimension $1/16$, which 
is the magnetic field along the preferential axis.\footnote{The conformal dimension of the magnetic field is fixed by the model.
It corresponds to the well-known critical exponent $1/8$ that governs 
the behavior of the spontaneous magnetization of the Ising model as the critical point is approached.}
This is exactly the perturbation that Zamolodchikov considered in his original paper.  
This means that if an ATFT is used to describe the magnetically perturbed Ising model,
we have no latitude in the choice of a Lie algebra: it must be $\E$. 

\subsection*{Why is it the compact form of $\E$?}
As Folland noted recently in \cite{Folland} physicists tend to think of Lie algebras in terms
of generators and relations, without even specifying a background field if they can help it.
So it can be difficult to judge from the appearance of a Lie algebra in the physics literature 
if any particular form of the algebra is being singled out.  

Nevertheless, the algebras appearing here are the compact ones.  
The reason is that the minimal model CFT's involve unitary representations of the Virasoro algebra.  The coset construction shows that these come from representations of affine Lie algebras which are themselves constructed from compact finite-dimensional algebras.  And it is these finite-dimensional Lie algebras that appear in the ATFT.

\subsection*{What about $\Eg6$ and $\Eg7$?}
So far, we have explained why it is $\E$ that is related to the cobalt niobate experiment.  This prompts the question: given a simple compact real Lie algebra $\g$, does it give a theory describing some other physical setup?  Or, to put it differently, what is the physical setup that corresponds to a theoretical model involving, say, $\Eg6$ or $\Eg7$?
 In fact, the field theories based on these other algebras do have interesting connections to statistical models.  For example, $\Eg7$ Toda field theory
describes the thermal perturbation of the tricritical Ising model, and the $\Eg6$ theory the thermal
deformation of the tricritical three-state Potts model.   These other models are easily
distinguished from the magnetically perturbed Ising model by their central charges. 
It will be interesting to see if physicists can come up with ways to probe these other
models experimentally.  The $\Eg7$ model might be easiest---the unperturbed, CFT version has already been realized, for example, in the form of helium atoms on krypton-plated graphite \cite{TFV}.

\section{The Zamolodchikov masses and $\E$'s publicity photo}

Translating Zamolodchikov's theory into the language of affine Toda field theory provides a way to transform his calculation of the particle masses listed in Table \ref{masses} into the solution of a rather easy system of linear equations, and that in turn is connected to the popular image of the $\E$ root system from Figure \ref{gosset.pop}.  These are connections that work for a general ATFT, and we write in that level of generality.

An ATFT is based on a compact semisimple real Lie algebra $\g$, such as the Lie algebra of the compact real $\E$.  We assume further that this algebra is simple and is not $\su(2)$.  Then from $\g$ we obtain 
a simple root system $R$ spanning $\R^\ell$ for some $\ell \ge 2$; this is canonically identified with the dual $\h^*$ of the Cartan subalgebra mentioned at the end of the previous section.  
 
We briefly explain how to make a picture like Figure \ref{gosset.actual} for $R$.  (For background
on the vocabulary used here,  please see \cite{Bou:g4} or \cite{Carter:simple}.)  
Pick a set $B$ of simple roots in $R$.  For each $\beta \in R$, write $s_\beta$ for the reflection in the hyperplane orthogonal to $\beta$.  The product $w := \prod_{\beta \in B} s_\beta$ with respect to any fixed ordering of $B$ is called a \emph{Coxeter element} and its characteristic polynomial has $m(x) := x^2 - 2 \cos(2\pi/h) x + 1$ as a simple factor \cite[VI.1.11, Prop.~30]{Bou:g4}, where $h$ is the Coxeter number of $R$.  The primary decomposition theorem gives a uniquely determined plane $P$ in $\R^\ell$ on which $w$ restricts to have minimal polynomial $m(x)$, i.e., is a rotation through $2\pi/h$---we call $P$ the \emph{Coxeter plane} for $w$.  The picture in Figure \ref{gosset.actual} is the image of $R$ under the orthogonal projection $\pi \!: \R^\ell \ra P$ in the case where $R = \E$.  We remark that while $P$ depends on the choice of $w$, all Coxeter elements are conjugate under the orthogonal group \cite[10.3.1]{Carter:simple}, so none of the geometric features of $\pi(R)$ are changed if we vary $w$ and we will refer to $P$ as simply a Coxeter plane for $R$.
 
In Figure \ref{gosset.actual}, the image of $R$ lies on 8 concentric circles.  This is a general feature of the projection in $P$ and is not special to the case $R = \E$.  Indeed, the action of $w$ partitions $R$ into $\ell$ orbits of $h$ elements each \cite[VI.1.11, Prop.~33(iv)]{Bou:g4}, and $w$ acts on $P$ as a rotation.  So the image of $R$ necessarily lies on $\ell$ circles.

The relationship between the circles in Figure \ref{gosset.actual} and physics is given by the following theorem.  
  
\begin{thm}\label{eq.thm}
Let $\g$ be a compact \emph{simple} Lie algebra that is not $\su(2)$,  and write $R$ for its root system.
For an affine Toda field theory constructed from $\g$, the following multisets are the same, up  to scaling by a positive real number:
\begin{enumerate}
\item \label{eq.mass} The (classical) masses of the particles in the affine Toda theory.
\item \label{eq.radii} The radii of the circles containing the projection of $R$ in a Coxeter plane.
\item \label{eq.PF} The entries in a Perron-Frobenius eigenvector for a Cartan matrix of $R$.
\end{enumerate}
\end{thm}

The terms in \eqref{eq.PF} may need some explanation.  The restriction of the inner product on $\R^\ell$ to $R$ is encoded by an $\ell$-by-$\ell$ integer matrix $C$, called the \emph{Cartan matrix} of $R$.  You can find the matrix for $R = \E$ in Figure \ref{cartan.fig}.  We know a lot about the Cartan matrix, no matter which $R$ one chooses---for example, its eigenvalues are all real and lie in the interval $(0, 4)$, see \cite[Th.~2]{BLM}.  Further, the matrix $2-C$ has all non-negative entries and is irreducible in the sense of the Perron-Frobenius Theorem, so its largest eigenvalue---hence the smallest eigenvalue of $C$---has a 1-dimensional eigenspace spanned by a vector $\vx$ with all positive entries.  (Such an eigenvector is exhibited in Figure \ref{cartan.PF} for the case $R = \E$.)  This $\vx$ is the vector in \eqref{eq.PF} and it is an eigenvector of $C$ with eigenvalue $4 \sin^2(\pi/2h)$, so calculating $\vx$ amounts to solving an easy system of linear equations.

\begin{figure}[hbt]
\centering
\subfloat[]{\label{cartan.fig}$\left( \begin{matrix}
2&0&-1&0&0&0&0&0 \\
0&2&0&-1&0&0&0&0 \\
-1&0&2&-1&0&0&0&0 \\
0&-1&-1&2&-1&0&0&0 \\
0&0&0&-1&2&-1&0&0 \\
0&0&0&0&-1&2&-1&0 \\
0&0&0&0&0&-1&2&-1 \\
0&0&0&0&0&0&-1&2
\end{matrix} \right)$}
\qquad
\subfloat[]{\label{cartan.PF}$\left( \begin{matrix}
m_2 \\
 m_4 \\ 
 m_6 \\
m_8\\ 
m_7 \\ 
m_5 \\ 
m_3\\
m_1
\end{matrix} \right) 
$}
\caption{The Cartan matrix (A) for the root system $\E$ and a Perron-Frobenius eigenvector (B), where the entries are as in Table \ref{masses}.}
\end{figure}

\begin{proof}[Sketch of proof] Theorem \ref{eq.thm} has been known to physicists since the early 1990s; here is a gloss of the literature.  Freeman showed that \eqref{eq.mass} and \eqref{eq.PF} are equivalent in \cite{Freeman}.  We omit his argument, which amounts to computations in the complex Lie algebra $\g \ot \C$, but it is worth noting that his proof does rely on $\g$ being compact.  

The equivalence of \eqref{eq.radii} and \eqref{eq.PF} can be proved entirely in the language of root systems and finite reflection groups, see for example \cite{FringLiaoOlive} or \cite[\S2]{Corrigan:recent}.  The Dynkin diagram (a graph with vertex set $B$) is a tree, so it has a 2-coloring $\s \!: B \ra \{ \pm 1 \}$, and one picks $w$ to be a corresponding Coxeter element as in \cite[\S10.4]{Carter:simple}.  Conveniently, the elements $\s(\beta) \beta$ for $\beta \in B$ are representatives of the orbits of $w$ on $R$, see \cite[p.250, (6.9.2)]{Kost:McKay} or \cite[p.~91]{FringLiaoOlive}.  It is elementary to find the inner products of $\pi(\s(\beta)\beta)$ with the basis vectors for $P$ given in \cite[\S10.4]{Carter:simple}, hence the radius of the circle containing $\pi(\s(\beta) \beta)$.  The entries of the Perron-Frobenius eigenvector appear naturally, because these entries are part of the expressions for the basis vectors for $P$.

Alternatively, Kostant shows the equivalence of \eqref{eq.mass} and \eqref{eq.radii} in \cite{Kost:gosset} using Lie algebras.
\end{proof}

There is a deeper connection between the particles in the ATFT and the roots in the root system.  Physicists identify the $w$-orbits in the root system with particles in the ATFT.  The rule for the coupling of particles in a scattering experiment (called a ``fusing" rule) is that the scattering amplitude for two particles $\O_1$ and $\O_2$ 
has a bound-state pole corresponding to $\O_3$ if and only if there are roots $\rho_i \in \O_i$ so that $\rho_1 + \rho_2 + \rho_3 = 0$ in $\R^\ell$, see \cite{Dorey:root} and \cite{FringLiaoOlive}. 
This leads to a ``Clebsch-Gordan" necessary condition for the coupling of particles, see \cite{Braden:note}.  We remark that these fusing rules are currently only theoretical---it is not clear how they could be tested experimentally.

\section{Back to the experiment}

Let's get back to the cobalt  niobate experiment.  As we noted above, 
when the external magnetic field is very close to the critical value
that induces the phase transition, it was expected that the experimental system would be modeled by the 
critical 1-dimensional quantum Ising model perturbed by a small magnetic field directed along the preferential axis.
This model is the subject of Zamolodchikov's perturbation theory,
and the resulting field theory has been identified as the $\E$ ATFT.

\begin{figure}[hbt]
\begin{center}
\subfloat[Masses detected]{\label{intensity.fig}\includegraphics[width=0.42\textwidth]{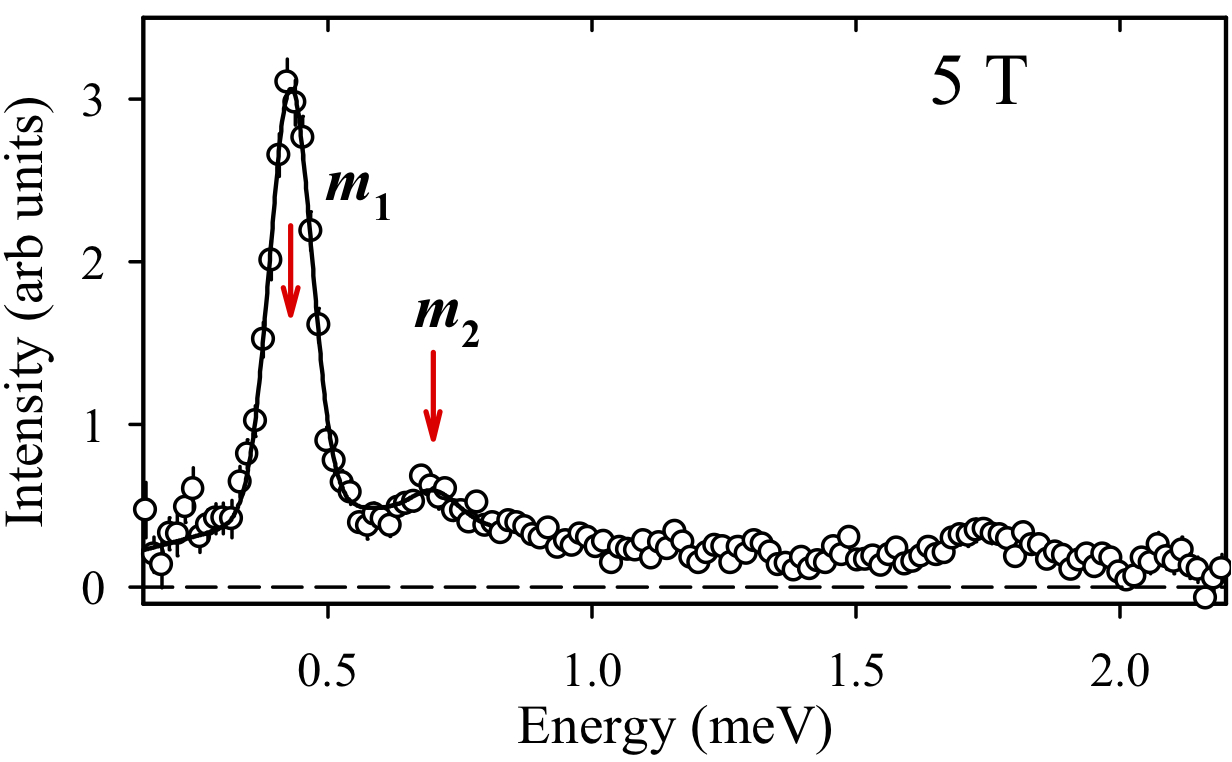}} 
\qquad
\subfloat[Predictions]{\label{predicted.fig}\raisebox{2.4ex}{\includegraphics[width= 
.3358\textwidth]{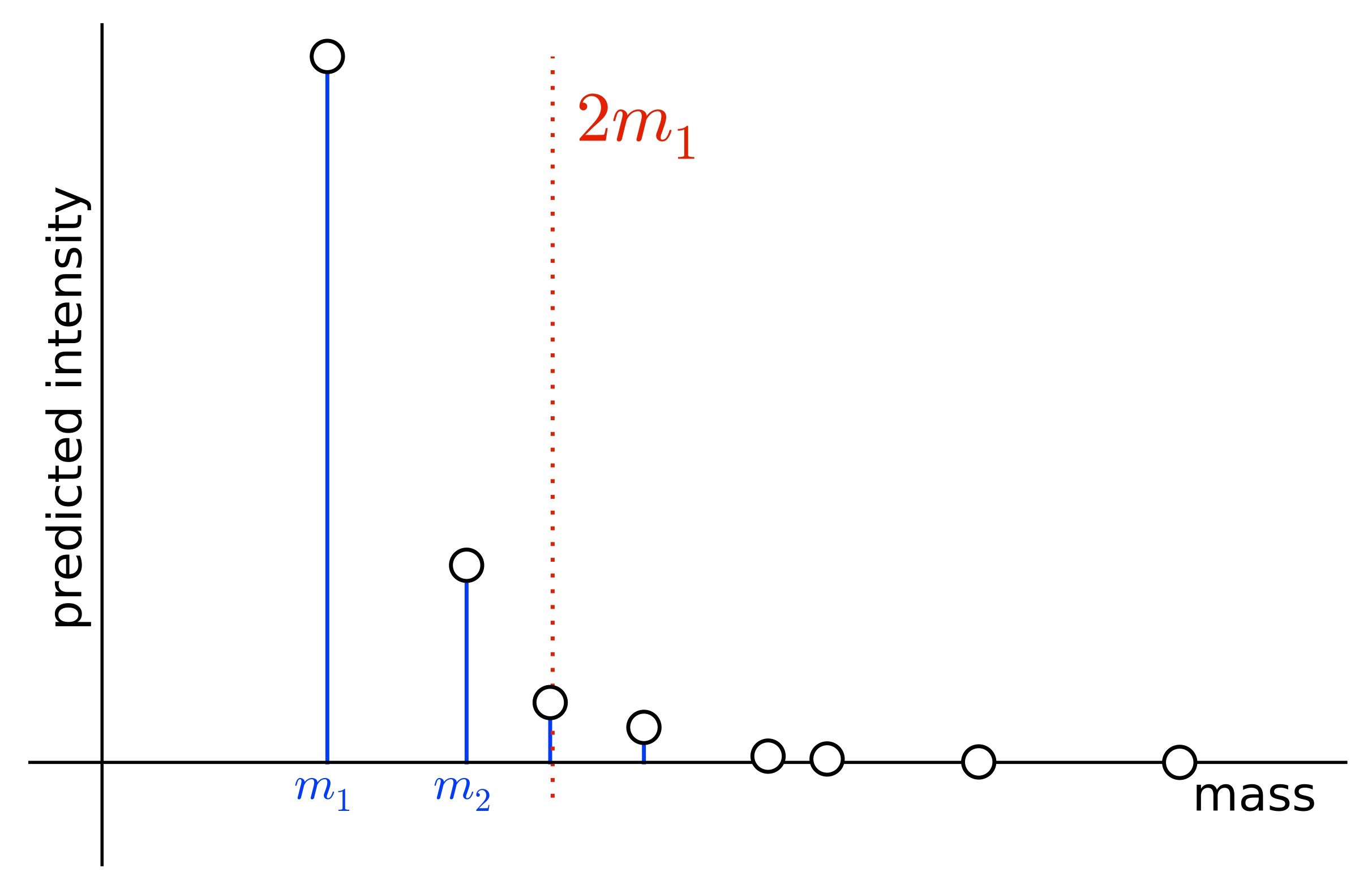}}} 
\caption{The left panel (A) is an example intensity plot, exhibiting the two detected masses under a transverse magnetic field of 5 tesla, 90\% of the critical strength.  \credit{Figure adapted from \cite{Coldea}.}  The right panel (B), shows the relative intensities obtained from the form factors computed in 
\cite[p.~741, Table 3]{DelfinoMussardo}.   The axes have the same labels as in the left panel.  The dotted vertical line marks the onset of the incoherent continuum.}\label{masspeaks.fig}
\end{center}
\end{figure}

To test this association, the experimenters conducted neutron scattering experiments
on the magnet.
Figure~\ref{intensity.fig} shows an intensity plot of scattered neutrons averaged over 
a range of scattering angles.  
Observations were actually made at a series of external field strengths, from 4.0 tesla (T) to 5.0 T,
with the second peak better resolved at the lower energies.  Both peaks track continuously
as the field strength is varied.   Figure~\ref{intensity.fig} represents the highest field strength
at which the second peak could be resolved.  

\begin{figure}[hbt]
\begin{center}
\includegraphics{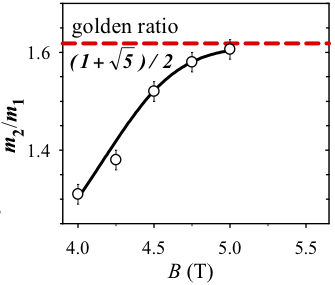}
\caption{The ratio $m_2/m_1$ of the masses of the two lightest particles approaches the golden ratio as the transverse magnetic field approaches critical strength of 5.5 tesla.   \credit{Figure adapted from \cite{Coldea}.} } \label{ratio.fig}
\end{center}
\end{figure}

The two peaks give evidence of the existence 
of at least two particles in the system, which was one of Zamolodchikov's core assumptions.  
And indeed, the ratio of the masses appears to approach the golden ratio---see Figure \ref{ratio.fig}---as the critical value (about 5.5 T) is approached, just as Zamolodchikov predicted twenty years earlier.

We can also compare the relative intensities of the first two mass peaks to the theoretical predictions exhibited in Figure \ref{predicted.fig}.  Here again we see approximate agreement between the observations
and theoretical predictions.    The figure shows a threshold 
at $2m_1$, where a continuous spectrum is generated 
by the scattering of the lightest particle with itself.  
Particles with masses at or above this threshold will be very difficult to detect, as their energy
signature is expected to consist of rather small peaks that overlap with the $2m_1$ continuum.\footnote{Possibly because of this, the region above this threshold has been called the ``incoherent continuum", a suggestive and Lovecraftian term.}   Hence the fact that only two particles out of eight were observed is again
consistent with the theoretical model.

\section{Experimental evidence for $\E$ symmetry?}

We can now finally address the question from the title of this paper, slightly rephrased: Did the experimenters detect $\E$?  First we should say that they themselves do not claim to have done so.  Rather, they claim to have found experimental evidence for the theory developed by Zamolodchikov, et al, and described above---which we shall call below simply \emph{Zamolodchikov's theory}---and that this in turn means giving evidence for $\E$ symmetry. 

The argument for these claims goes as follows.  The $\E$ ATFT is an integrable field theory describing the magnetically perturbed Ising model \eqref{Hmag} and satisfying \eqref{Z0}--\eqref{Z3}.  In that situation, Zamolodchikov and Delfino-Mussardo made some numerical predictions regarding the relative masses 
of the particles and relative intensities of the scattering peaks.   The experimental data show
two peaks, but the second peak is only resolved at lower energies.  The ratios of masses and intensities
are certainly consistent with the theoretical predictions, although the ratios appear to be measured only 
rather roughly.


At this point, we want to address three objections to this line of argument that we heard when giving talks on the subject.

\subsection*{Objection \#1: confirmatory experimental results are not evidence} We heard the following objection: experiments can never provide evidence for a scientific theory, they can only provide evidence against it.  (This viewpoint is known as \emph{falsificationism}.)  This is of course preposterous.  Science only progresses through the acceptance of theories that have survived enough good experimental
tests, even if the words ``enough" and ``good" are open to subjective interpretations.  

A less extreme version of this same objection is: confirmatory experimental results are automatically suspect in view of notorious historical examples of experimenter's bias such as cold fusion and N-rays.  This sort of objection is better addressed to the experimental physics community, which as a whole is certainly familiar with these specific examples and with the general issue of experimenter's bias.  As far as we know, no such criticisms have been raised concerning the methods described in \cite{Coldea}.

\subsection*{Objection \#2: it still doesn't seem like enough data} Recall that the experimental results can be summarized as a limited set of numbers which approximately agree with the theoretical predictions.  Based on this, we have heard the following objection: if you start by looking at this small amount of data, how can you claim to have pinned down something as complex as $\E$?  This question contains its own answer.  One doesn't analyze the results of the experiment by examining the data, divorced from all previous experience and theoretical framework.  Instead, humanity already knows a lot about so-called critical point phenomena\footnote{See for example the 20 volume series \emph{Phase transitions and critical phenomena} edited by C.~Domb and J.L.~Lebowitz.} and there is a substantial theoretical model that is expected to describe the behavior of the magnet.  The experiment described in \cite{Coldea} was a test of the relevance and accuracy of Zamolodchikov's theory, not an investigation of magnets beginning from no knowledge at all.

To put it another way, someone who approaches science from the viewpoint of this objector would necessarily reject many results from experimental physics that are based on similar sorts of indirect evidence.  To give just one example of such a result, the reported observations of the top quark in \cite{D0} and \cite{CDF} were not direct observations but rather confirmations of theoretical predictions made under the assumption that the top quark exists.

\subsection*{Objection \#3: the numerical predictions don't require $\E$}
If you examine the papers \cite{Zam:int} by Zamolodchikov and \cite{DelfinoMussardo} by Delfino and Mussardo, you see that the numerical predictions are made without invoking $\E$.  At this point, one might object that $\E$ is not strictly necessary for the theoretical model.  But,
as we explained in \S\ref{Zam.sec}, the role of $\E$ in the theory is that by employing it, Zamolodchikov's \emph{assumption} \eqref{Z0} is turned into a \emph{deduction}.  That is, by including $\E$, we reduce the number of assumptions and achieve a more concise theoretical model.  Moreover, the $\E$ version of the theory justifies the amazing numerological coincidences 
between Zamolodchikov's calculations and the $\E$ root system.

\subsection*{Evidence for $\E$ symmetry?}
Finally, we should address the distinction between ``detecting $\E$" and ``finding evidence for $\E$ symmetry".  While the former is pithier, we're only talking about the latter here.  The reason is that, as far as we know, there is no direct correspondence between $\E$ and any physical object.  This is in contrast, for example, to the case of the gauge group $\SU(3)$ of the strong force in the Standard Model in particle physics.  One can meaningfully identify basis vectors of the Lie algebra $\su(3)$ with gluons, the mediators of the strong force, which have been observed in the laboratory.
With this distinction in mind, our view is that the experiment cannot be said to have detected $\E$, but that it has provided evidence for Zamolodchikov's theory and hence for $\E$ symmetry as claimed in \cite{Coldea}. 

\section{Summary}

The experiment with the cobalt niobate magnet consisted of two phases.  In the first phase, the experimenters verified that in the absence of an external magnetic field,
the 1-dimensional quantum Ising model \eqref{Hmag} accurately describes the spin dynamics, as predicted by theorists.  In the second phase, the experimenters added an external magnetic field directed transverse to the spins' preferred axis,
and tuned this field close to the value required to reach the quantum critical regime.  
In that situation, Zamolodchikov, et al, had predicted the existence of 8 distinct types of particles in a field theory governed by the compact Lie algebra $\E$.  The experimenters observed the two smallest particles and confirmed two numerical predictions: the ratio of the masses of the two smallest particles (predicted by Zamolodchikov) and the ratio of the intensities corresponding to those two particles (predicted by Delfino-Mussardo).  

In this article, we have focused on the $\E$ side of the story because $\E$ is a mathematical celebrity.  But there is a serious scientific reason to be interested in the experiment apart from $\E$: it is the first experimental test of the perturbed conformal field theory constructed by Zamolodchikov around 1990.  Also, it is the first laboratory realization of the critical state of the quantum
1-dimensional Ising model in such a way that it can be manipulated---the experimenters can continuously vary the transverse field strength $g_x$ in \eqref{Hmag} across a wide range while preserving the 1-dimensional character---and the results observed directly.  
Since the Ising model is the fundamental
model for quantum phase transitions, the opportunity to probe experimentally its very rich physics
represents a breakthrough.


\newcommand{\etalchar}[1]{$^{#1}$}
\providecommand{\bysame}{\leavevmode\hbox to3em{\hrulefill}\thinspace}
\providecommand{\MR}{\relax\ifhmode\unskip\space\fi MR }
\providecommand{\MRhref}[2]{%
  \href{http://www.ams.org/mathscinet-getitem?mr=#1}{#2}
}
\providecommand{\href}[2]{#2}

\end{document}